\documentstyle[osa,manuscript,graphicx,epsf]{revtex}  

\begin{document}                
\newcommand{\beq}{\begin{equation}} 
\newcommand{\eeq}{\end{equation}}
\newcommand{\beqa}{\begin{eqnarray}}
\newcommand{\eeqa}{\end{eqnarray}}
\title{SUSY Scaling Violations and UHECR}

\author{Claudio Corian\`{o}\footnote{presented by C. Corian\`{o} at the Intl. 
Workshop ``QCD @ work'', Martina Franca, Italy 16-20 June 2001} }
\address{Dipartimento di Fisica Universita' di Lecce\\
 I.N.F.N. Sezione di Lecce Via Arnesano, 73100 Lecce, Italy}
\author{Alon E. Faraggi}
\address{Theoretical Physics Department,
University of Oxford, Oxford, OX1 3NP, United Kingdom\\
and Theory Division, CERN, CH--1211 Geneva, Switzerland}

\maketitle
\vspace{1cm}
\centerline{ \em Dedicated to the Memory of Prof. Nathan Isgur}
\begin{abstract}
Advancing QCD toward astroparticle 
applications generates new challenges for perturbation theory, such as  
the presence of large evolution scales with sizeable scaling violations 
involving both the initial and the final state of a collision. 
Possible applications in the context of 
Ultra High Energy Cosmic Rays (UHECR) of these effects are discussed. 
\end{abstract}

\section{ INTRODUCTION}
Nowadays intriguing theoretical extensions of the Standard Model 
are being explored, while the experimental results continue to confirm the
validity of the model and constrain its extensions.
Given the limited energy range available at colliders, it is therefore
vital to develop experimental probes that can climb up the energy ladder.
A complementary way to analize these extensions while waiting for 
colliders of the next generations is provided by cosmic rays, which exceed
the energy scales currently attainable.

We have undertaken the preliminary steps in a program
that aims to utilize cosmic rays as an experimental probe of 
theoretical generalizations of the Standard Model. 
We have in mind possible applications of cosmic rays for the study of 
supersymmetry. The first is in the context of top-down models of
Ultra--High--Energy Cosmic Rays (UHECR), around and above the
Greisen--Zatsepin--Kuzmin cutoff. 
In these models the primary cosmic rays originate from the
decay of a metastable superheavy particle which decay at rest,
fragmenting into ordinary hadrons and photons. 
The dynamics of these decays can be modelled using standard QCD tools 
on which we elaborate below. 
We propose to analize supersymmetric effects in the 
decay of these metastable states using 2 scales:
\begin{itemize}
\item[]
A High Energy $\Lambda_F$ fragmentation Scale $\approx 10^{11}$ 
(decay of a metastable state $\rightarrow$ primary protons) 
\item[]
A collision scale $\Lambda_{coll}$ due to the
interaction of surviving primaries with air-nuclei 
($E_{\rm CoM}\approx 10-400$ TeV)
\end{itemize}
At both scales supersymmetric scaling violations 
should be included and the multiplicities of the spectrum analized. 

\section{Numerical Results}
As an illustration of the procedure we adopt in our studies, let's consider the decay 
of a hypothetical massive state of mass 1 TeV into supersymmetric partons. 
The decay can proceed, for instance, through a regular $q\bar{q}$ 
channel and a shower is developed starting from the quark pair. The $N=1$ DGLAP 
equation describes in the leading logarithmic approximation the evolution 
of the shower which accompanies the pair, and we are interested in studying the 
impact of the supersymmetry breaking scale ($m_\lambda$) on the fragmentation. 
In our runs we have chosen the initial set of Ref.~\cite{kkp}.          

We parameterize the fragmentation functions as
\begin{equation}
\label{temp}
D(x,\mu^2)=Nx^\alpha(1-x)^\beta\left(1+\frac{\gamma}{x}\right)
\end{equation}

Typical fragmentation functions in QCD involve final states with 
$p$, $\bar{p}$, $\pi^{\pm},\pi^{0}$ and kaons $k^{\pm}$.
 We have chosen 
an initial evolution scale of $10$ GeV and varied both the mass of the SUSY partners 
(we assume for simplicity that these are all degenerate) 
and the final evolution scale. 
In general the effects of supersymmetric evolution 
are small within the range described by the factorization scales $Q_f$ and 
$Q_i$ ($Q_f=10^3$ GeV, $Q_i=200$ GeV). 
We mention that $Q_f$ is the starting scale (the highest scale) 
at which the decay of the supersymmetric partons starts. $Q_i$ 
is fixed by the gluino/squark masses and coincides with them. 

The situation appears to be completely different 
for the gluon fragmentation functions (f.f's) (Fig.~1). The regular and the SQCD evolved 
f.f.'s differ largely in the diffractive region, and this clearly will show up in the 
spectrum of the primary protons if the decaying state has a supersymmetric content. 
As we raise the final evolution scale 
we start seeing more pronounced differences between regular 
and supersymmetric distributions. 
We have shown in Fig.~2 the squark f.f.'s for all the flavours and 
the one of the gluino for comparison. The scalar charm distribution appear 
to grow slightly faster then the remaining scalar ones. The gluino f.f. 
is still the fastest growing at small-x values.

\section{Summary}
In a few years several experiments, including the Pierre Auger experiment
\cite{auger}, will start collecting data
from cosmic rays. The issue of the origin of UHECR will
be -hopefully- clarified.  
While the link of UHECR to AGN's has been disfavored on the basis of a
quite homogeneous distributions,
the local origin of these events remains an open possibility.
Potential meta--stable superheavy string relics have been suggested as
dark matter candidates, as well as potential
sources for the UHECR \cite{CFP}.
A QCD/SQCD analysis of these events is in progress.
With the forthcoming experimental data \cite{auger}, and improved 
theoretical analysis, along the lines discussed here, cosmic 
ray physics enters an exciting new era, with potentially
ground--breaking discoveries.

\begin{figure}[htb]
\centerline{\epsfxsize 3.0 truein \epsfbox{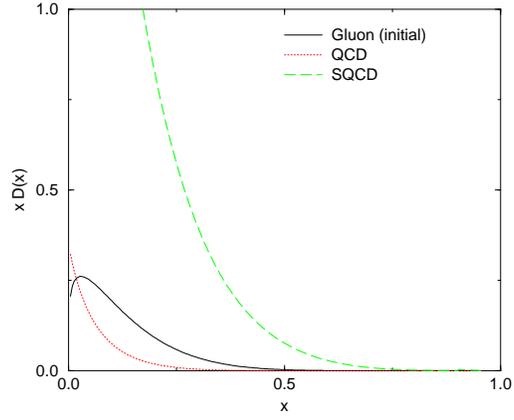}}
\caption[]{\small 
The gluon fragmentation function $x D_{g}^{p,\bar{p}}(x,Q^2)$  at the lowest scale (input)
$Q_0=10$ GeV, and its evolved QCD (regular) and SQCD/QCD evolutions 
with $Q_f=$$10^3$ GeV.The SUSY fragmentation scale is chosen to be $200$ GeV.}
\end{figure}

\begin{figure}[ht]
\centerline{\epsfxsize 3.0 truein \epsfbox{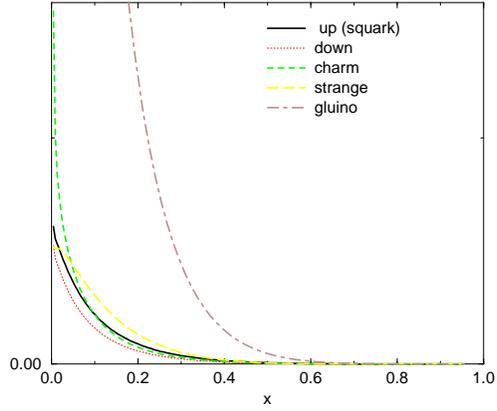}}
\caption[]{\small The fragmentation functions of squarks and gluino
 at the lowest scale (input)
$Q_0=10$ GeV, with $Q_f=10^3$ GeV and SUSY scale $200$ GeV.}
\end{figure}

\begin{figure}
\centerline{\includegraphics[angle=-90,width=.4\textwidth]{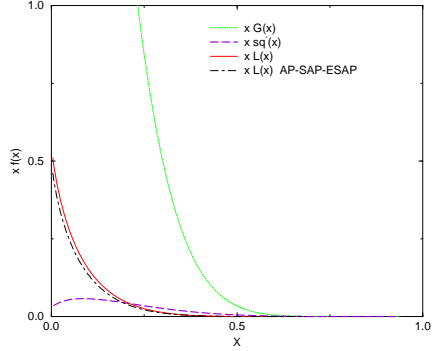}}
\caption{{ $x f(x)$ in the AP-ESAP evolution with a very large final scale 
$Q_f=10^3$ GeV and with 
a squark mass $m_{\tilde{q}}=100$ GeV. Shown are the non-singlet squark, 
the gluon and the gluino distributions for the AP-ESAP evolution. The gluino distribution for the AP-SAP-ESAP evolution is also shown 
(with $m_{2\lambda}=40$ GeV). }}
\label{quark15}
\end{figure}
\bigskip 
{\bf Acknowledgments:} A.F. Thanks the CERN theory division for hospitality.
The work of C.C. is supported in part by INFN 
(iniziativa specifica BARI-21) and by MURST. The work of A.F. is supported
by PPARC.

\end{document}